\documentclass[oneside, reqno, 11pt]{amsart}

\usepackage[a4paper, margin=2.7cm]{geometry}

\usepackage{amsmath}
\usepackage{amsfonts}
\usepackage{graphicx}
\usepackage[dvipsnames]{xcolor}

\usepackage{mathtools}
\usepackage{amsthm}

\newtheorem{theorem}{Theorem}[section]

\newtheorem{remark}[theorem]{Remark}
\newtheorem{defn}[theorem]{Definition}

\usepackage{url}
\usepackage[hyperfootnotes=false, colorlinks, citecolor=RoyalBlue, urlcolor=blue, linkcolor=blue ]{hyperref}

\usepackage{placeins}

\AtBeginDocument{%
  }

\usepackage[linesnumbered, ruled, vlined, algosection]{algorithm2e}
\SetKwInput{KwComplexity}{Complexity}
\SetKwInput{KwRuleI}{Rule~I}
\SetKw{RuleI}{Rule~I}
\SetKwInput{KwRuleII}{Rule~II}
\SetKw{RuleII}{Rule~II}
\SetKw{RuleIandII}{Rules~I\,\&\,II}
\SetKw{Continue}{continue}
\SetKw{Break}{break}
\SetKw{False}{false}
\SetKw{True}{true}
\SetKw{And}{and}
\SetKw{Or}{or}
\SetKw{Delete}{delete}
\SetKw{Not}{not}
\SetKw{Then}{then}

\usepackage{caption}
\usepackage{tabularray}
\UseTblrLibrary{siunitx}
\UseTblrLibrary{counter}
\UseTblrLibrary{functional}
\UseTblrLibrary{diagbox}

\DefTblrTemplate{caption-tag}{caption-ams-style}{\textsc{Table~\thetable .}}
\DefTblrTemplate{caption-sep}{caption-ams-style}{\enskip}
\DefTblrTemplate{caption-text}{caption-ams-style}{\InsertTblrText{caption}}
\DefTblrTemplate{conthead-text}{caption-ams-style}{(Continued~on~next~page)}

\ExplSyntaxOn
\DefTblrTemplate{caption}{caption-ams-style}{\vspace{0.7em}
	\hbox_set:Nn \l__tblr_caption_box
	{
		\UseTblrTemplate { caption-tag } { default }
		\UseTblrTemplate { caption-sep } { default }
		\UseTblrTemplate { caption-text } { default }
	}
	\dim_compare:nNnTF { \box_wd:N \l__tblr_caption_box } > { 0.85\linewidth }
	{
		\noindent
		\makebox[\linewidth][c]{
			\parbox{0.85\textwidth}{
				\hbox_unpack:N \l__tblr_caption_box
			}
		}
		\par
	}
	{
		\centering
		\makebox [\linewidth] [c] { \box_use:N \l__tblr_caption_box }
		\par
	}
}
\DefTblrTemplate{capcont}{caption-ams-style}{\vspace{0.7em}
	\hbox_set:Nn \l__tblr_caption_box
	{
		\UseTblrTemplate { caption-tag } { default }
		\UseTblrTemplate { caption-sep } { default }
		\UseTblrTemplate { caption-text } { default }
		\UseTblrTemplate { conthead-pre } { default }
		\UseTblrTemplate { conthead-text } { default }
	}
	\dim_compare:nNnTF { \box_wd:N \l__tblr_caption_box } > { 0.85\linewidth }
	{
		\noindent
		\makebox[\linewidth][c]{
			\parbox{0.85\textwidth}{
				\hbox_unpack:N \l__tblr_caption_box
			}
		}
		\par
	}
	{
		\centering
		\makebox [\linewidth] [c] { \box_use:N \l__tblr_caption_box }
		\par
	}
}
\ExplSyntaxOff

\DefTblrTemplate{firsthead,middlehead,lasthead}{caption-bottom-ams-style}{
}
\DefTblrTemplate{firstfoot}{caption-bottom-ams-style}{
	\UseTblrTemplate{capcont}{caption-ams-style}
}
\DefTblrTemplate{middlefoot}{caption-bottom-ams-style}{
	\UseTblrTemplate{capcont}{caption-ams-style}
}
\ExplSyntaxOn
\DefTblrTemplate{lastfoot}{caption-bottom-ams-style}{
	\UseTblrTemplate{note}{default}
	\UseTblrTemplate{remark}{default}
		\UseTblrTemplate{caption}{caption-ams-style}
}
\ExplSyntaxOff

\NewTblrTheme{ams-theme}{
	\SetTblrTemplate{caption-tag}{caption-ams-style}
	\SetTblrTemplate{caption-sep}{caption-ams-style}
	\SetTblrTemplate{caption-text}{caption-ams-style}
	
	\SetTblrTemplate{conthead-text}{caption-ams-style}
	\SetTblrTemplate{caption}{caption-ams-style}
	
	\SetTblrTemplate{firsthead,middlehead,lasthead,firstfoot,middlefoot,lastfoot}{caption-bottom-ams-style}
}

\DefTblrTemplate{caption-tag}{caption-IEEE-style}{TABLE~\thetable}
\DefTblrTemplate{caption-sep}{caption-IEEE-style}{}
\DefTblrTemplate{caption-text}{caption-IEEE-style}{\InsertTblrText{caption}}
\DefTblrTemplate{conthead-pre}{caption-IEEE-style}{ }
\DefTblrTemplate{conthead-text}{caption-IEEE-style}{(Continued)}
\DefTblrTemplate{contfoot-text}{caption-IEEE-style}{Continued on next column/page.}

\ExplSyntaxOn
\DefTblrTemplate{caption}{caption-IEEE-style}{
	{
		\noindent
		\makebox[\hsize][c]{
			\parbox{\hsize}{
				\centering
				\footnotesize
				\textsc{
				\UseTblrTemplate { caption-tag } { caption-IEEE-style } \\
				\UseTblrTemplate { caption-text } { caption-IEEE-style }
				}
			}
		}
		\par
	}
}
\DefTblrTemplate{capcont}{caption-IEEE-style}{\vspace{0.0em}
	{
		\noindent
		\makebox[\hsize][c]{
			\parbox{\hsize}{
				\centering
				\footnotesize
				\textsc{
					\UseTblrTemplate { caption-tag } { caption-IEEE-style } \\
					\UseTblrTemplate { caption-text } { caption-IEEE-style }
					\UseTblrTemplate { conthead-pre } { caption-IEEE-style }
					\UseTblrTemplate { conthead-text } { caption-IEEE-style }
				}
			}
		}
		\par
	}
}
\ExplSyntaxOff

\NewTblrTheme{IEEE-theme}{
	\SetTblrTemplate{caption-tag}{caption-IEEE-style}
	\SetTblrTemplate{caption-sep}{caption-IEEE-style}
	\SetTblrTemplate{caption-text}{caption-IEEE-style}
	\SetTblrTemplate{contfoot-text}{caption-IEEE-style}
	
	\SetTblrTemplate{caption}{caption-IEEE-style}
	\SetTblrTemplate{capcont}{caption-IEEE-style}
	
}

\DefTblrTemplate{caption-tag}{caption-acm-style}{\textbf{Table~\thetable :}}
\DefTblrTemplate{caption-sep}{caption-acm-style}{ }
\DefTblrTemplate{caption-text}{caption-acm-style}{\textbf{\InsertTblrText{caption}}}
\DefTblrTemplate{conthead-text}{caption-acm-style}{(Continued)}

\ExplSyntaxOn
\DefTblrTemplate{caption}{caption-acm-style}{\vspace{0.0em}
	\hbox_set:Nn \l__tblr_caption_box
	{
		\UseTblrTemplate { caption-tag } { default }
		\UseTblrTemplate { caption-sep } { default }
		\UseTblrTemplate { caption-text } { default }
	}
	\dim_compare:nNnTF { \box_wd:N \l__tblr_caption_box } > { \hsize }
	{
		\noindent
		\makebox[\hsize][c]{
			\parbox{\hsize}{
				\hbox_unpack:N \l__tblr_caption_box
			}
		}
		\par
	}
	{
		\centering
		\makebox [\hsize] [c] { \box_use:N \l__tblr_caption_box }
		\par
	}
}
\DefTblrTemplate{capcont}{caption-acm-style}{\vspace{0.0em}
	\hbox_set:Nn \l__tblr_caption_box
	{
		\UseTblrTemplate { caption-tag } { default }
		\UseTblrTemplate { caption-sep } { default }
		\UseTblrTemplate { caption-text } { default }
		\UseTblrTemplate { conthead-pre } { default }
		\UseTblrTemplate { conthead-text } { default }
	}
	\dim_compare:nNnTF { \box_wd:N \l__tblr_caption_box } > { \hsize }
	{
		\noindent
		\makebox[\hsize][c]{
			\parbox{\hsize}{
				\hbox_unpack:N \l__tblr_caption_box
			}
		}
		\par
	}
	{
		\centering
		\makebox [\hsize] [c] { \box_use:N \l__tblr_caption_box }
		\par
	}
}
\ExplSyntaxOff

\DefTblrTemplate{firsthead,middlehead,lasthead}{caption-bottom-acm-style}{
}
\DefTblrTemplate{firstfoot}{caption-bottom-acm-style}{
}
\DefTblrTemplate{middlefoot}{caption-bottom-acm-style}{
}
\ExplSyntaxOn
\DefTblrTemplate{lastfoot}{caption-bottom-acm-style}{
	\UseTblrTemplate{note}{default}
	\UseTblrTemplate{remark}{default}
	\UseTblrTemplate{caption}{caption-acm-style}
}
\ExplSyntaxOff

\NewTblrTheme{acm-theme}{
	\SetTblrTemplate{caption-tag}{caption-acm-style}
	\SetTblrTemplate{caption-sep}{caption-acm-style}
	\SetTblrTemplate{caption-text}{caption-acm-style}
	
	\SetTblrTemplate{conthead-text}{caption-acm-style}
	\SetTblrTemplate{caption}{caption-acm-style}
	
	\SetTblrTemplate{firsthead,middlehead,lasthead,firstfoot,middlefoot,lastfoot}{caption-bottom-acm-style}
}

\usepackage{chngcntr}
\numberwithin{equation}{section}

\usepackage{enumitem}
\usepackage{appendix}

\usepackage{rotating}

\usepackage{fewerfloatpages}

\newcommand{\sm}{\left(\begin{smallmatrix}} \newcommand{\esm}{\end{smallmatrix}\right)} \newcommand{\bpm}{\begin{pmatrix}} \newcommand{\ebpm}{\end{pmatrix}}

\newcommand{\vect}[1]{{\boldsymbol{#1}}}

\newcommand{\rquotient}[2]{
	\mathchoice
	{
		\text{\raise01ex\hbox{$#1$}\Big/\lower1ex\hbox{$#2$}}%
	}
	{
		#1\,/\,#2
	}
	{
		#1\,/\,#2
	}
	{
		#1\,/\,#2
	}
}

\newcommand{\bigrquotient}[2]{
	\mathchoice
	{
		\text{\raise01ex\hbox{$#1$}\Bigg/\lower1ex\hbox{$#2$}}%
	}
	{
		#1\,/\,#2
	}
	{
		#1\,/\,#2
	}
	{
		#1\,/\,#2
	}
}

\newcommand{\cpp}{{C\nolinebreak[4]\hspace{-.35em}\raisebox{.12ex}{ {+\hspace{-.0em}+}}}}

\counterwithin{table}{section}
\counterwithin{figure}{section}

\begin{document}

\title{Elasticity in Parallel Sparse Triangular Solve}


\author{Raphael S. Steiner}
\address{Huawei, Thurgauerstrasse 80, 8050 Zurich, CH}
\email{raphael.steiner@huawei.com}

\author{Christos K. Matzoros}
\address{Huawei, Thurgauerstrasse 80, 8050 Zurich, CH}
\email{christos.konstantinos.matzoros@h-partners.com}

\author{P\'al Andr\'as Papp}
\address{Huawei, Thurgauerstrasse 80, 8050 Zurich, CH}
\email{pal.andras.papp@huawei.com}

\author{Toni B\"ohnlein}
\address{Huawei, Thurgauerstrasse 80, 8050 Zurich, CH}
\email{toni.boehnlein@huawei.com}

\author{Albert-Jan N. Yzelman}
\address{Huawei, Thurgauerstrasse 80, 8050 Zurich, CH}
\email{albert-jan@yzelman.net}

\begin{abstract} We introduce stale synchronous parallel as a mode of execution in parallel sparse triangular linear system solve and present a general directed-acyclic-graph scheduler capable of producing such schedules.
Stale-synchronous-parallel schedules allow the overlap of synchronisation and compute which results in a geometric-mean speed-up of $7$-$30\%$ of our scheduler, ElasticDivide, over state-of-the-art synchronous scheduler GrowLocal on an ARM machine using 48 cores. On an x86 machine using 48 cores, we report geometric-mean speed-ups of $19$-$60\%$ over SpMP.
\end{abstract}

\keywords{%
	Sparse triangular linear system solve, SpTrSV, SpTrSM, forward- and backward-substitution algorithm, stale-synchronous-parallel algorithm.
}

\maketitle

%
%
%
\vspace{-5mm}

\section{Introduction}
\label{sec:intro}

Sparse triangular solve is an omnipresent operation in computing. Whether it be in engineering, data analytics, artificial intelligence, or scientific computing, systems of equations need to be solved.
This typically involves (sparse) triangular solve as part of the solving process directly, through methods like LU, QR, and Cholesky factorisations and Gau{\ss}--Seidel, or indirectly through (pre-)conditioning of the system for faster iterative solving methods such as conjugate gradient.

When solving ever bigger problem instances, sparsity plays an evermore important role.
It allows one to cut redundant compute, but this comes at the cost of losing structure.
This results in fine-grained dependencies and operations which makes sparse triangular solve (SpTrSV) a hard problem to parallelise.
Therefore, getting the most performance out of modern multi-core architectures is a challenging endeavour.

To combat this, SpTrSV performance is often optimised in larger contexts such as linear or symmetric solves where preprocessing allows for the reintroduction of structure.
A prominent example of this is the nested dissection technique \cite{NestedDissectionAlan, GeneralisedND, karypis1998fast, PermBlockDiagonalForm, Hund, mondriaan} which concentrates non-zeroes for locality and introduces coarse-grained parallelism \cite{BlockdiagonalSparseSolvers, UnifieedCommOptimisationSpTrSV}.
Another important technique is the introduction of so-called supernodes \cite{supernodes1, SuperLU, PanguLU, supernodes2, supernodes3, PardisoUnsymmetric, PardisoSymmetric}.
Supernodes give rise to several optimisations. First, the grouping of operations into supernodes allows for the use of highly optimised dense kernels, and second, the dependency graph becomes coarser which enables dynamically computing a parallel schedule or dynamic dispatching methods as their benefits now outweigh their overhead.
Another notable and generally applicable technique is the (recursive) splitting of SpTrSV into two SpTrSV problems of half the size and an easier-to-parallelise SpMV in-between \cite{anderson1989solving, mayer2009parallel, liu2016synchronization, lu2020efficient, ahmad2021split}.

After applying these structural techniques (if they are feasible), one is left with a parallel scheduling problem.
The tasks to be scheduled typically have irregular dependencies and their size can range from whole blocks or supernodes to a couple of floating-point operations.
It is this scheduling problem that we address in this paper.
More precisely, we consider the fine-grained scheduling problem of parallelising the forward-substitution algorithm, cf.\@ Algorithm~\ref{alg:forward-substitution}.
Nevertheless, our algorithm and the involved ideas may, of course, be applied to any parallel scheduling problem.

\begin{figure*}[!t]
	\centering
	\includegraphics[width=0.98\linewidth]{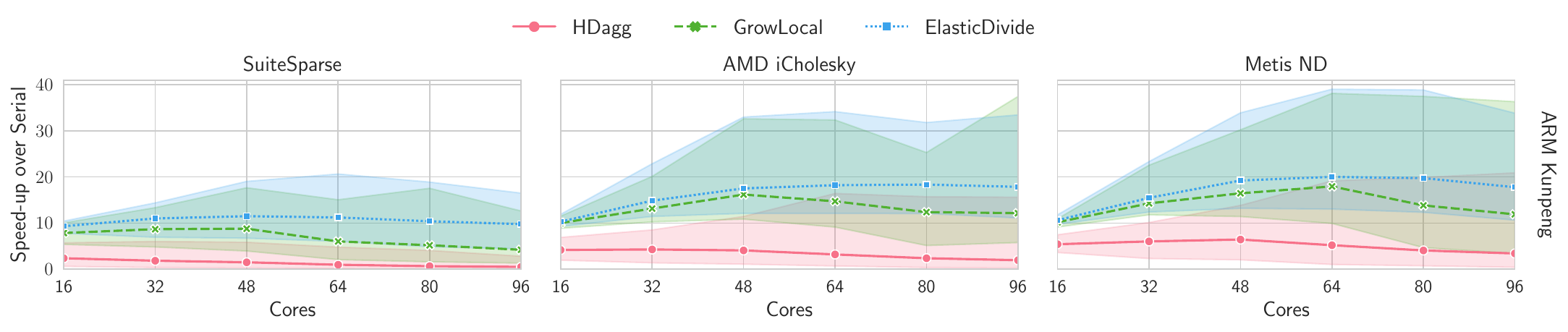}
	\caption{\label{fig:intro-scaling} Geometric-mean speed-ups over Serial for various data sets, cf.\@~\S\ref{sec:data-sets}, on ARM Kunpeng with 10-th to 90-th percentile range shown.}
\end{figure*}

To get the most performance out of SpTrSV, a parallel scheduling algorithm must
\begin{itemize}
	\item balance workloads among cores,
	\item limit coordination overhead, and
	\item consider spatial and temporal locality.
\end{itemize}
Several algorithms to this end have been proposed in the literature.
Broadly speaking, they come in two flavours, synchronous and asynchronous, depending on the type of schedule they produce.
Early algorithms include asynchronous self-scheduling~\cite{saltz1988run} and the synchronous so-called wavefront schedulers~\cite{anderson1989solving, saltz1990aggregation}.
Both approaches suffered from a lot of synchronisation overhead due to frequent synchronisation~\cite{rothberg1992parallel, park2014sparsifying}.
In a breakthrough paper~\cite{park2014sparsifying}, Park \emph{et al.\@} addressed this issue.
Their scheduler SpMP combines the coarseness from the grouping of wavefront schedulers with asynchronous compute and point-to-point synchronisation.
To further reduce the number of synchronisations, they employed an approximate transitive reduction to get rid of the majority of implied (and thus unnecessary) dependencies.
Y{\i}lmaz \emph{et al.\@}~\cite{yilmaz2020adaptive} were further able to reduce the number of dependencies by forcing the cores to be out-of-sync by only a bounded amount.
Additionally, their scheduler ALB takes into account the amount of parallelism available locally in the matrix as the computation moves forward to decide how many of the available cores to use.

On the synchronous front, there have also been improvements.
The HDagg scheduler by Zarebavani \emph{et al.\@}~\cite{zarebavani2022hdagg} combines successive wavefronts whenever beneficial in order to reduce the number of synchronisations.
In contrast, B\"ohnlein \emph{et al.\@}~\cite{GrowLocalIPDPS, GrowLocalArXiv} moved away from wavefronts and took inspiration from list scheduling~\cite{graham1969bounds, adam1974comparison, hwang1989scheduling, radulescu2002low, mingsheng2003efficient}.
Their scheduler, GrowLocal, interleaves local list scheduling with synchronisation barriers which communicate the completion of rows.
This allowed for much longer independent compute, reducing the number of synchronisations by an order of magnitude.

\subsection{Our contribution}

In this paper, we strike a middle ground between synchronous and asynchronous compute.
Namely, we employ for the first time stale synchronous parallel~\cite{StaleSyncPaper, StaleSyncPaperProperDef} as the mode of execution in sparse triangular solve.
This execution model explicitly overlaps synchronisation and compute by forcing inter-core dependencies to be elongated.
In other words, the information computed by one core can only be used on another core after several synchronisation events.
Since the information is not immediately required after a synchronisation event, the synchronisation can be overlapped with other compute based on already locally available information.
In the case of synchronisation-heavy workloads with many threads, we show that this can result in a $2\times$ geometric-mean speed-up over synchronous execution.
On architectures with higher number of NUMA-domains, this can go up to $4.5\times$.
Even when the architecture is more uniform and the synchronisation events are few, it leads to improvement in the double digit percentages.

Naturally, stale synchronous parallel puts additional strain on the schedule to be generated and therefore also on the scheduler.
Nevertheless, we demonstrate empirically that the computational graphs associated to an SpTrSV of a matrix are elastic enough to allow for such schedules.
Indeed, our scheduler \emph{ElasticDivide} produces about the same number of synchronisation barriers ($0.85\times$ to $1.28\times$) as the state-of-the-art synchronous scheduler GrowLocal~\cite{GrowLocalArXiv} when appropriately normalised, yet overlaps compute with synchronisation.
In Figure~\ref{fig:intro-scaling}, we depict geometric-mean speed-ups over Serial on an ARM Kunpeng machine with 48 cores and two NUMA-domains per socket.
The three plots correspond to three data sets which are derived from the SuiteSparse Matrix Collection~\cite{davis2011university}.
SuiteSparse consists of matrices directly taken from the data set, and the other two data sets are application oriented with the matrices having been preprocessed with an approximate-minimal-degree ordering and incomplete Cholesky using Eigen~\cite{eigenweb}, respectively with nested dissection using Metis~\cite{karypis1998fast}.
The reported geometric-mean speed-ups range from $3\%$ on small core counts to over $100\%$ on large core counts over the second best measured algorithm GrowLocal.
On 48 cores (single socket), the geometric-mean speed-ups are $30\%$, $7\%$, and $17\%$, respectively.
SpMP, being x86-specific, is missing from Figure~\ref{fig:intro-scaling}.
On 48 cores on an AMD EPYC machine, we achieve geometric-mean speed-ups of $19\%$, $60\%$, and $26\%$, respectively, over SpMP with geometric-mean speed-ups reaching over $100\%$ on smaller core counts.

In summary, in this paper, we
\begin{itemize}
	\item propose and analyse solving SpTrSV problems via stale synchronous parallelisation,
	\item introduce ElasticDivide, a stale-synchronous-parallel scheduler for general directed acyclic graphs, and
	\item show their effectiveness on non-uniform memory architectures.
\end{itemize}

\subsection{Overview}
\label{sec:overview}
We describe the stale-synchronous-parallel scheduling problem in~\S\ref{sec:prelim}. The ElasticDivide scheduler is presented in detail in~\S\ref{sec:algorithms}. In~\S\ref{sec:implementation}, we describe the implementation of the kernel and the barrier. In~\S\ref{sec:experiment-setup} and~\S\ref{sec:evaluation}, we describe our evaluation procedure and evaluate the algorithms, followed by a discussion in~\S\ref{sec:conclusion}.

\ifdefined\anonymous \else
\subsection*{Acknowledgements}
We would like to thank Weifeng Liu, Olaf Schenk, Xiaoye Li, Dimosthenis Pasadakis, Lorenzo Migliari, and Kiril Dichev for stimulating conversations on this and surrounding topics.

\fi

\FloatBarrier

\section{Preliminaries}
\label{sec:prelim}

\subsection{Problem description}
\label{sec:problem-description}

In sparse triangular solve (vector), one is given a sparse invertible matrix $\vect{A} = (A_{i,j})_{i,j = 1, \dots, n} \in \mathbb{R}^{n \times n}$ and a vector $\vect{b} = (b_i)_{i = 1, \dots, n} \in \mathbb{R}^n$ and seeks to compute the vector $\vect{x} = (x_i)_{i = 1, \dots, n} \in \mathbb{R}^n$ such that
\begin{equation}
	\vect{A}\vect{x} = \vect{b}
\end{equation}
holds. The standard algorithm to solve for $\vect{x}$ is the \emph{for\-ward-sub\-sti\-tu\-tion algorithm} if the matrix $\vect{A}$ is lower triangular, respectively the \emph{back\-ward-sub\-sti\-tu\-tion algorithm} if $\vect{A}$ is upper triangular. Assume from now on that $\vect{A}$ is lower triangular. The forward-substitution algorithm is described in Algorithm~\ref{alg:forward-substitution}.

\begin{algorithm}[htbp]
	\DontPrintSemicolon
	\SetNlSty{textsc}{}{}
	\SetAlgoNlRelativeSize{-1}
	\caption{Forward-substitution algorithm\label{alg:forward-substitution}}
	\KwData{An invertible lower-triangular matrix $\vect{A}\in \mathbb{R}^{n \times n}$ and a vector $\vect{b} \in \mathbb{R}^{n}$.}
	\KwResult{A vector $\vect{x}\in \mathbb{R}^{n}$ such that $\vect{A}\vect{x}=\vect{b}$.}
	\BlankLine
	\For{$i = 1, \dots, n$\label{alg-line:forward-sub-outer-loop}}{
		$t \gets b_i$\;
		\For{\emph{$j = 1, \dots, i-1$ with $A_{i,j}\neq 0$}}{
			$t \gets t - A_{i,j}x_j$\;
		}
		$x_i \gets t / A_{i,i}$\;
	}
\end{algorithm}

In contrast to the dense case, when the matrix $\vect{A}$ is sparse, there is opportunity to parallelise the outer for-loop, cf.\@ Line~\ref{alg-line:forward-sub-outer-loop}. Though, one has to be careful as there are dependencies. The $i$-th iteration depends directly on the $j$-th iteration if and only if $A_{i,j} \neq 0$ and $i \neq j$. More generally, the $i$-th iteration depends on the $j$-th iteration if and only if there is a sequence $i = \ell_0 > \ell_1 > \dots > \ell_m = j$ with $A_{\ell_k, \ell_{k+1}} \neq 0$ for $k = 0, \dots, m-1$ for some $m \ge 1$. We can and shall capture this in a directed acyclic graph $G_{\vect{A}}=(V_{\vect{A}}, E_{\vect{A}})$, where
\begin{align}
	V_{\vect{A}} &= \{1, \dots, n\}, \ \text{and} \\
	E_{\vect{A}} &= \{(j,i) \in V_{\vect{A}} \times V_{\vect{A}} \mid A_{i,j} \neq 0 \}\,.
\end{align}
We may also attach a (vertex-)weight to $G_{\vect{A}}$ modelling the execution time of the $i$-th iteration corresponding to a vertex $i\in V_{\vect{A}}$.
A natural choice as this weight is the number of non-zero elements the corresponding row, that is
\begin{equation}
	\begin{aligned}
		\omega_{\vect{A}}: V_{\vect{A}} &\to \mathbb{Z}_{> 0} \\
		i &\mapsto \sharp\{j \in V_{\vect{A}} \mid A_{i,j} \neq 0 \}\,.
	\end{aligned}
\end{equation}
This is also equal to the in-degree plus one in the graph representation.

From here on out, we work with the graph representation $G=G_{\vect{A}}$ of the sparse lower triangular matrix $\vect{A}$ and formulate the problem as a scheduling problem in an execution model described in the next section.

\subsection{Stale synchronous parallel}
\label{sec:stale-synchronous-parallel}

Stale synchronous parallel is a computational model extending the bulk-synchronous-parallel model \cite{valiant1990bridging, valiant1990general}. It was formally introduced by Cui \emph{et al.\@}~\cite{StaleSyncPaperProperDef} and originally stems from a paper by Cipar \emph{et al.\@}~\cite{StaleSyncPaper} in machine learning. Similar to the bulk synchronous parallel, stale synchronous parallel operates in `rounds' which are called \emph{supersteps}, but, unlike bulk synchronous parallel, cores may move the computation ahead into subsequent supersteps as long as the maximal difference in the superstep index remains bounded by a fixed quantity called the \emph{staleness}. Formally, a compute schedule in the stale-synchronous-parallel model may be defined as follows.

\begin{defn} \label{defn:stale-synchronous-parallel-schedule}
	A \emph{stale-synchronous-parallel schedule} of \emph{staleness} $s \in \mathbb{Z}_{>0}$ of a directed acyclic graph $G=(V,E)$ to a set of cores $P$ consists of a pair $(\pi, \sigma)$ of maps $\pi:V \to P$ and $\sigma:V \to \mathbb{Z}_{\ge 0}$ satisfying
	\begin{equation}
		\forall (v,w) \in E:\ \sigma(v) + s \cdot \delta_{\pi(v),\pi(w)} \le \sigma(w)\, ,
	\end{equation}
	where $\delta_{i,j}$ is the Kronecker delta. The map $\pi$ maps vertices to cores and the map $\sigma$ maps vertices to \emph{supersteps}.
\end{defn}

\begin{remark} \label{rem:bsp-schedule-is-ssp-schedule}
	The definition of a stale-synchronous-parallel schedule of staleness $s=1$ is the same as a bulk-synchronous-parallel schedule.
\end{remark}

\begin{remark} \label{rem:ssp-schedule-also-smaller-staleness}
	A stale-syn\-chro\-nous-par\-al\-lel schedule of staleness~$s$ is naturally a stale-syn\-chro\-nous-par\-al\-lel schedule of staleness~$t$ such that $1 \le t \le s$, in particular also a bulk-synchronous-parallel schedule.
\end{remark}

The use of the word `stale' in the name `stale synchronous parallel' is perhaps unfortunate and stems from the first paper~\cite{StaleSyncPaper}, where it was used in machine-learning model training. Workers were allowed to compute model updates based on a mildly outdated (stale) model, which allowed for some asynchronicity whilst (provably) maintaining convergence and quality guarantees. Only later~\cite{StaleSyncPaperProperDef}, it was realised that the `data' need not be outdated (or stale), but instead having `long' dependencies suffices.

The stale-synchronous-parallel model has since found many applications, such as recommendation systems~\cite{KirilRecommend}, matrix factorisation~\cite{SspApplicationsForMl}, topic modeling~\cite{SspApplicationsForMl}, and training for deep learning~\cite{SspInDeepLearning}. 
We also note that overlapping computation and communication, a key motivation for the stale synchronous parallel model, had already been identified as a core feature of the bulk synchronous parallel model~\cite{valiant1990bridging}.

\section{The ElasticDivide scheduler}
\label{sec:algorithms}

Our scheduling algorithm ElasticDivide is an extension of the algorithm \emph{GrowLocal} presented in~\cite{GrowLocalArXiv}.
A sketch of ElasticDivide is given by Algorithm~\ref{alg:GrowLocalSSP}. The main novelty of our algorithm is the adaptation to a stale-synchronous-parallel schedule of staleness~$2$, cf.\@ Definition~\ref{defn:stale-synchronous-parallel-schedule}.
To achieve this, we had to innovate and add two key features:

\begin{enumerate}[label=(\roman*)]
	\item a delay in resolving dependencies across cores, and
	\item strike a balance between assigning vertices in the current superstep and the next.
\end{enumerate}

\begin{algorithm}[htbp]
	\DontPrintSemicolon
	\SetNlSty{textsc}{}{}
	\SetAlgoNlRelativeSize{-1}
	\caption{Sketch of ElasticDivide scheduler\label{alg:GrowLocalSSP}}
	\KwData{A vertex-weighted directed acyclic graph $G = (V,E,\omega)$ and cores $P = \{1,2,\dots,k\}$.}
	\KwResult{Assignments of vertices to cores $\pi:V \to P$ and superstep $\sigma: V \to \mathbb{Z}_{\ge 0}$.}
	\BlankLine
	\KwRuleI{A vertex $v$ is \emph{assignable} to core $p$ and superstep $s \in \mathbb{Z}_{\ge 0}$ if and only if $$\forall(w,v)\in E: \sigma(w)+2\cdot \delta_{\pi(w),p} \le s\,.$$}
	\KwRuleII{Vertices are prioritised according to
		\begin{enumerate}[label=(\roman*),leftmargin=1.94cm, rightmargin=0.5cm, topsep=.5pt]
			\item core exclusivity, and then
			\item smallest ID.
	\end{enumerate}}
	
	\BlankLine
	$s \gets 0$\;
	\While{\emph{not all vertices are assigned}}{
		$\alpha \leftarrow 20$ \;
		\While{\True\!\!\label{alg-line:superstep-attempt}}{
			\tcp{I. Assign vertices to cores}
			assign up to $\alpha$ \emph{assignable} vertices to core $1$ and superstep $s$ with \RuleIandII\;
			$\Omega_1 \leftarrow$ total newly assigned weight to core $1$\;
			\For{{\rm core} $p=2,\dots, k$ {\rm in} {\rm order}}{
				$\Omega_p \leftarrow 0$\;
				\While{$\Omega_p \not \approx \Omega_1$\label{alg-line:GrowLocal-weight-balance}}{
					\uIf{\emph{$\exists$ \emph{assignable} vertex $v$ to core $p$ and superstep $s$}}{
						assign vertex $v$ to core $p$ and superstep $s$ with \RuleIandII\;
						$\Omega_p \leftarrow \Omega_p + \omega(v)$\;
					}{
						\lElse{\Break}	
					}
				}
			}
			
			\tcp{II. Score assignments}
			
			$\displaystyle \beta \leftarrow \frac{\sum_p \Omega_p}{\max_p \Omega_p + 2000}$ \label{alg-line:parallelisation-score}\;
			
			${\rm exit} \gets \True$\;
			\If{\emph{the score $\beta$ is high enough}}{
				deem current assignments as worthy\;
				${\rm exit} \gets \False$\;
			}
			\If{\emph{not enough \emph{assignable} vertices for next superstep}\label{alg-line:ssp-future-superstep-ready}}{
				${\rm exit} \gets \True$
			}
			undo vertex assignments with superstep $s$\;
			$\alpha \leftarrow \tfrac{3}{2} \alpha$\;
			
			\If{${\rm exit}$}{
				redo last worthy assignments\;
				$s \gets s+1$\;
				\Break inner loop\;
			}
		}
	}
\end{algorithm}

The algorithm proceeds by assigning vertices in the order of supersteps. During each superstep it assigns \emph{assignable} vertices to cores, cf.\@ \emph{Rule~I}.
Rather than going through time steps to keep a work balance between cores, as you would in (barrier) list schedulers \cite{graham1969bounds,adam1974comparison, hwang1989scheduling, radulescu2002low, mingsheng2003efficient,Papp2024Efficient}, the algorithm has a guess $\alpha$ as to how long it may compute during a superstep and fills up each core up to that limit in order.
This allows the algorithm to prioritise locality of the execution according to \emph{Rule~II}.
The second nature of \emph{Rule~I} is that it extends compute during a single superstep by assigning more restricted vertices first.
Once no further assignments are being made, the assignments in their totality are measured by a scoring function, cf.\@ Line~\ref{alg-line:parallelisation-score}.
If the score is high enough, the assignments are considered successful and another attempt with increased $\alpha$ is initiated.
The increase of $\alpha$ is exponential as to amortise the costs of retrying assignments of the current superstep.
Starting with a small $\alpha$ and gradually increasing it allows the algorithm to keep a work balance, but also maximise independent compute. The balance between work balance and independent compute is determined by the score $\beta$.
The scoring function is taken from \cite{GrowLocalArXiv} and is inversely proportional to the execution time of the current superstep in a bulk-synchronous-parallel model without communication \cite{valiant1990bridging} scaled up to the whole graph.
The constant $2000$ thus plays the role of the synchronisation cost.
Albeit our algorithm produces a stale-synchronous-parallel schedule with staleness $2$ rather than a bulk-synchronous-parallel schedule and can thus overlap synchronisation and compute, we find that the scoring function nevertheless does a good job in balancing parallelism and the length of independent compute.

Returning to the flow of the algorithm, the superstep attempts are interrupted when either the score $\beta$ starts to drop or it is deemed that there are no longer sufficient \emph{assignable} vertices left for the following superstep.
At this point, the best assignment measured by the score $\beta$ is chosen as the final assignment of the current superstep and the algorithm moves onto the next superstep.

The algorithm is efficiently implemented using sorted double-ended queues for the globally ready queues of vertices which are \emph{assignable} to all cores for a given superstep, and heap for locally ready queues of vertices which are \emph{assignable} to a \emph{single} given core and given superstep.
As a result, the presented algorithm runs in almost linear time under reasonable assumptions. We refer to \cite[Appendix~B]{GrowLocalArXiv} for a proof and \S\ref{sec:amortisation} for empirical data.

At last, we also remark on the choice of staleness~$2$. It is the smallest number that allows the overlap of compute and synchronisation. Larger numbers would only be beneficial if the compute is too small compared to the synchronisation. We find this to be unlikely the case. Moreover, in our initial testing, we have found that larger staleness negatively affects the schedule quality, thus degrading the overall performance. Hence, we chose staleness equal to~$2$.

\section{SpTrSV implementation}
\label{sec:implementation}

\subsection{Synchronisation mechanism}
\label{sec:barrier}

A stale-synchronous-parallel schedule of staleness $s \ge 2$ allows one to overlap synchronisation with computation. To this end, we implemented a weak (or fuzzy) barrier~\cite{fuzzy-barrier} which separates the {\tt arrive} and {\tt wait} operations.

In our barrier implementation, each core has its own flag to signal that it has arrived at the barrier. 
The flag-owning core has exclusive write access to its flag. 
All other cores may read this flag to check whether the flag-owning core has arrived at the barrier. 
This allows us to avoid read-modify-write operations. 
In addition, our barrier makes use of caching \cite{SPSCRingBufferCaching, rigtopSPSCQueue}, that is, every core has their own local cache of the flags of all other cores. This reduces contention and communication.

To make the most out of the caching mechanism, to save memory, and to reduce code complexity, we combine all barriers into one by having a counter as the flag. 
This counter records the superstep which the core is currently processing or is waiting to start processing, meaning the core has completed the computation of all previous supersteps. 
This greatly benefits the caching mechanism as the flag of a core which is (far) ahead in terms of supersteps only needs to be checked once until the probing core has caught up. 
In fact, the probing core not only can catch up but also move ahead of the cached value due to the staleness before having to read again the flag of the same core.

The {\tt wait} operation was implemented using busy waiting.

For other kinds of barriers, we refer to the survey paper~\cite{hoefler2005survey} and the references therein.

\subsection{Stale-synchronous-parallel SpTrSV kernel} \label{sec:kernel}

Our stale-synchronous-parallel SpTrSV kernel solves a lower triangular linear system in parallel according to a stale-synchronous-parallel schedule $(\pi, \sigma)$ of staleness $2$, cf.\@ \S\ref{sec:stale-synchronous-parallel}.
We parallelise the kernel using OpenMP~\cite{dagum1998openmp} threads.
Each thread runs on its own physical core and represents a core $p$ in the schedule.
In each superstep $s$, every core $p$ executes the computations associated to the rows whose corresponding vertex $v$ has been assigned to said processor and superstep, i.e.\@, $v \in \pi^{-1}(\{p\})\cap \sigma^{-1}(\{s\})$, in ascending order.
The supersteps are surrounded with the weak barrier mechanics from \S\ref{sec:barrier}, that is, the \texttt{wait} operation at the beginning of the superstep and \texttt{arrive} at the end of the superstep.
We note that the \texttt{wait} operation only needs to be invoked if there is something to compute for the core during the superstep.
If there is not, the \texttt{wait} can be safely skipped.
As supersteps are sometimes empty due matrix-parallelism limitations and algorithm choices, this leads to improvements due to the implemented caching mechanism of the barrier, cf.\@ \S\ref{sec:barrier}.

\begin{algorithm}[htbp]
	\DontPrintSemicolon
	\SetNlSty{textsc}{}{}
	\SetAlgoNlRelativeSize{-1}
	\caption{\label{alg:ssp-sptrsv} Stale-synchronous-parallel SpTrSV kernel}
	\KwData{An invertible lower-triangular matrix $\vect{A} \in \mathbb{R}^{n \times n}$, a vector $\vect{b}\in \mathbb{R}^n$, a set of cores $P=\{1,2,\dots,k\}$, and a stale-synchronous-parallel schedule $(\pi,\sigma)$ for $G_{\vect{A}}$ of staleness $2$, cf.\@ \S\ref{sec:stale-synchronous-parallel}.}
	\KwResult{A vector $\vect{x}\in \mathbb{R}^n$ such that $\vect{A}\vect{x}=\vect{b}$.}
	\BlankLine
	\ForPar{\emph{core} $p \in P$}{
		\For{$s = 0,1,\dots, \max \sigma(V_{\vect{A}})$}{
			\lIf{$\pi^{-1}(\{p\}) \cap \sigma^{-1}(\{s\}) \neq \emptyset$}{
				\texttt{wait}$(p, s - 2)$
			}
			\For{$i \in \pi^{-1}(\{p\}) \cap \sigma^{-1}(\{s\})$}{
				$t \leftarrow b_i$\;
				\For{$j = 1, \dots, i - 1$ \emph{with} $A_{i,j} \neq 0$}{
					$t \leftarrow t - A_{i,j} x_j$\;
				}
				$x_i \leftarrow t / A_{i,i}$\;
			}
			\texttt{arrive}$(p, s)$\;
		}
	}
\end{algorithm}

\section{Experimental setup}
\label{sec:experiment-setup}

\begin{figure*}[!htbp]
	\centering
	\includegraphics[width=0.97\linewidth]{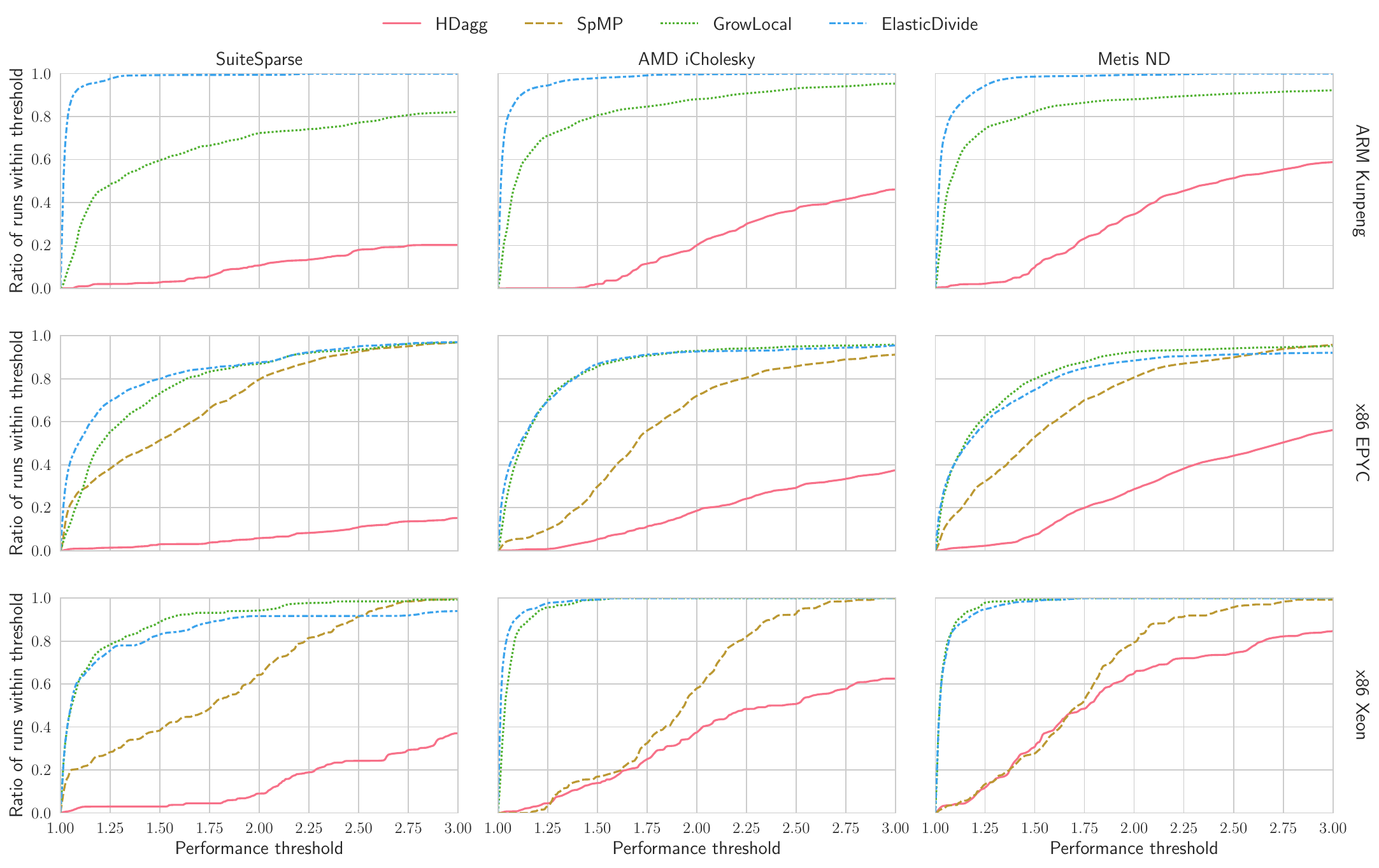}
	\caption{\label{fig:performance-overall} Performance profiles of the various SpTrSV scheduling algorithms on a given data set and CPU architecture. In each plot, the x-axis represents a threshold and the y-axis the ratio of runs of a given algorithm which are within this threshold times the best run of the same matrix and core count.}
\end{figure*}

This section presents the benchmarks and setup used for our experiments. In large part, we follow the experimental setup of~\cite{GrowLocalIPDPS, GrowLocalArXiv} and~\cite{zarebavani2022hdagg}.

\ifdefined\anonymous
The implementation of our algorithm and the test suite is open source on GitHub, with the link hidden for the sake of double blind reviewing.
\else
The implementation of our algorithm and the test suite are available open source within the \emph{OneStopParallel} scheduling framework on Github~\cite{OneStopParallel}.
\fi

\subsection{Methodology}
Our own scheduling algorithm is evaluated using the SpTrSV kernel described in \S\ref{sec:kernel}. The kernel is parallelised through the OpenMP library~\cite{dagum1998openmp}, with \texttt{OMP\_PROC\_BIND}=\texttt{close} and \texttt{OMP\_PLACES}=\texttt{cores} as the settings. The scheduler GrowLocal is implemented in a similar fashion with the appropriate bulk-synchronous-parallel kernel implementation.

The baseline schedulers HDagg and SpMP are evaluated using the Sympiler framework of Chesmi \emph{et al.}~\cite{cheshmi2017sympiler, cheshmi2022transforming}. This includes the original implementation of HDagg~\cite{zarebavani2022hdagg} and has SpMP~\cite{park2014sparsifying} already integrated into it.

For every scheduler and problem instance, we repeat the SpTrSV execution $100$ times in order to obtain a reliable estimation of the running time. We also add two untimed executions beforehand in order to ensure that the measurements all happen in a `hot' state of the system. Between two consecutive SpTrSV executions, the right-hand-side vector $\vect{b}$ is always reset to the all-ones vector.

The test suite of our kernel inherently implements the process above. With minimal modifications, we also adjusted the runtime measurement module of Sympiler to follow the same methodology. 

All scheduling algorithms (ours and the baselines) are implemented in \cpp~and were compiled with GCC (11.4.0 or 11.5.0) using the optimisation flag \texttt{-O3}. The running times of the algorithms are measured using the high-resolution clock in \cpp~(available in \texttt{std::chrono}).

\subsection{Data sets}
\label{sec:data-sets}

We evaluate the SpTrSV scheduling algorithms on several different data sets.
The data sets all originate from the SuiteSparse Matrix Collection \cite{davis2011university}. This collection contains a diverse set of matrices from a variety of applications, and it is the prominent evaluation benchmark in previous works on SpTrSV scheduling~\cite{park2014sparsifying,zarebavani2022hdagg,GrowLocalIPDPS, GrowLocalArXiv}.
Following in large part~\cite{GrowLocalIPDPS, GrowLocalArXiv} and~\cite{zarebavani2022hdagg}, one of our data sets consists of matrices directly from SuiteSparse. The other two data sets consist of modified versions of the same matrices in order to align more closely with applications and previous benchmarks.
An overview of the matrices in the data sets as well as some basic information, including the number of non-zero entries and the average wavefront size, may be found in \S\ref{sec:tables-matrices}.

\subsubsection{SuiteSparse} \label{sec:florida-graphs}

This data set is taken from~\cite{GrowLocalIPDPS, GrowLocalArXiv} and consists of the lower triangular part of real symmetric positive definite matrices from the SuiteSparse Matrix Collection~\cite{davis2011university}. 
It encompasses $33$ matrices. The SpTrSV problem associated with each matrix involves at least $2$ million floating point operations and has an average wavefront size of at least $44$.
An overview of the properties of these matrices is available in Table~\ref{table:florida-graph-table} of the supplement.

\subsubsection{SuiteSparse Eigen incomplete Cholesky (iChol)} \label{sec:cholesky-graphs}

Our second data set is also taken from~\cite{GrowLocalIPDPS, GrowLocalArXiv} and consists of lower triangular matrices obtained from an incomplete Cholesky decomposition on the real symmetric matrices from the SuiteSparse data set, see \S\ref{sec:florida-graphs}.
The incomplete Cholesky decomposition was performed using the `IncompleteCholesky' method in Eigen \cite{eigenweb}, which entailed applying its internal reordering method `AMD\-Ordering'~\cite{AMDOrdering}. We note that the matrix `bundle\_adj' segmentation-faults during this process.
The properties of the matrices in this data set are outlined in Table~\ref{table:cholesky-graph-table} of the supplement.

\subsubsection{SuiteSparse METIS (METIS)} \label{sec:metis-graphs}

For the final data set, we use the data set considered in~\cite{zarebavani2022hdagg}. It is once more derived from real symmetric positive definite matrices from the SuiteSparse Matrix Collection. The matrices are preprocessed using the fill-reducing method (nested dissection) of the METIS partitioner~\cite{karypis1998fast}, which symmetrically permutes the original matrices. After this preprocessing step, the lower triangular part is taken.

The matrices in this data set are representative of SpTrSV workloads in a Gau{\ss}--Seidel or a zero-fill-in incomplete Cholesky preconditioned conjugate gradient method for sparse symmetric solve.
An overview of the matrices in this data set is available in Table~\ref{table:hdagg-metis-graph-table} of the supplement.

\subsection{CPU architectures}
\label{sec:cpu-architectures}

We evaluated our SpTrSV scheduler on both x86 and ARM architectures. The concrete specifications for the three machines used in the experiments are as follows:

\begin{itemize}
	\item Huawei Kunpeng 920-4826 (Hi1620) processor (ARM, introduced 2019), with 512~GB memory, theoretical peak memory throughput of 187.7~GB/s, and two sockets with 48 cores each for a total of 96 cores; 2 NUMA domains per socket; kernel version 5.15.0; GCC version 11.4.0.
	\smallbreak
	\item AMD EPYC 7763 processor (x86, introduced 2021), with 1024~GB memory, theoretical peak memory throughput of 204.8~GB/s, and two sockets with 64 cores each for a total of 128 cores; 1 NUMA domain per socket; kernel version 5.15.0; GCC version 11.4.0;
	\smallbreak
	\item Intel Xeon Gold 6238T processor (x86, introduced 2019), with 192~GB memory, theoretical peak memory throughput of 140.8~GB/s, and two sockets with 22 cores each for a total of 44 cores; 1 NUMA domain per socket; kernel version 5.14.0; GCC version 11.5.0; 
\end{itemize}

\section{Evaluation}
\label{sec:evaluation}

\begin{figure*}[!htbp]
	\centering
	\includegraphics[width=0.97\linewidth]{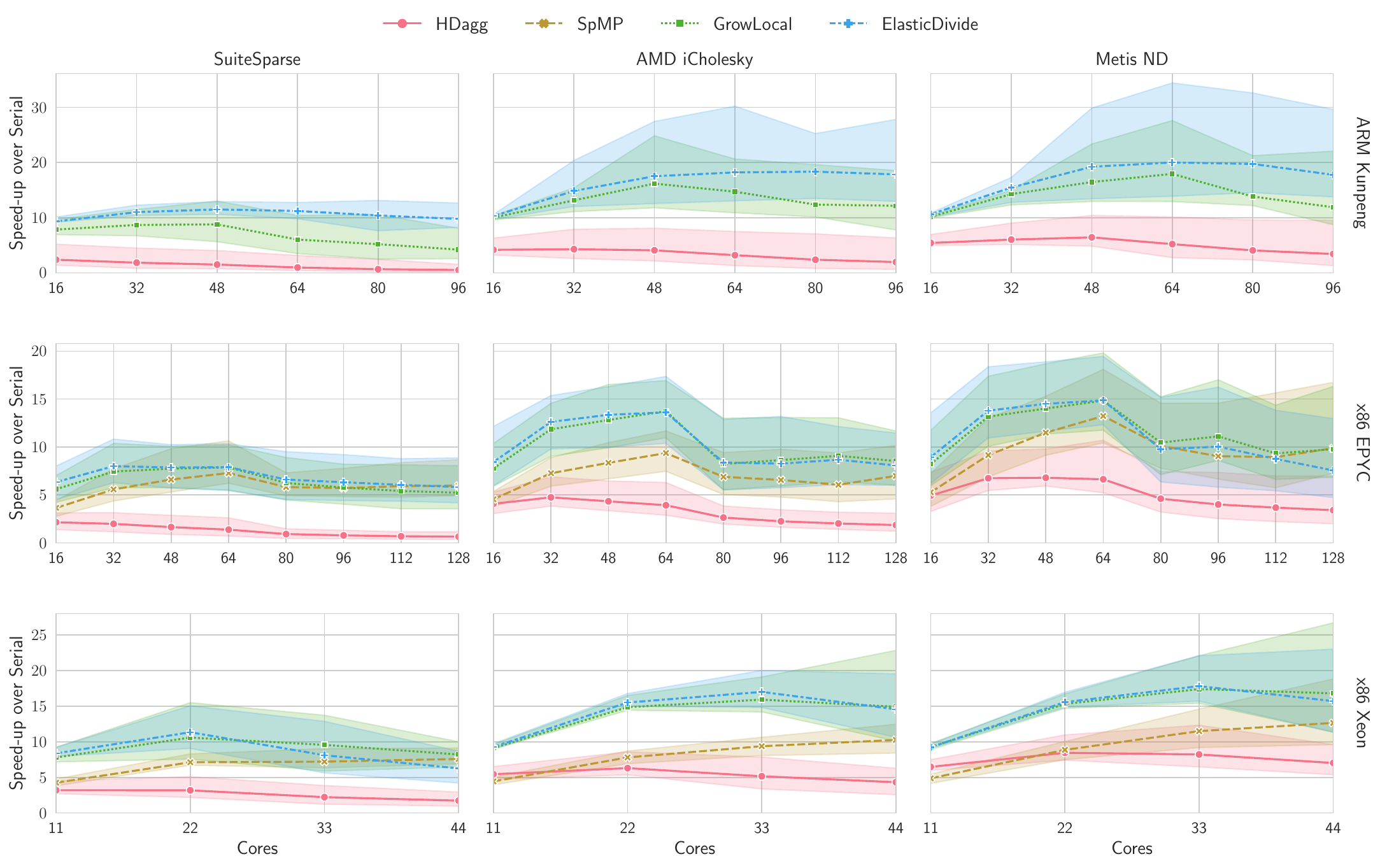}
	\caption{\label{fig:scaling} Scaling plots depicting the geometric mean speed-up over Serial and interquartile range of each algorithm on a given data set, architecture, and core count.}
\end{figure*}

\subsection{Runtime}

To summarise our findings, we have condensed in Figure~\ref{fig:performance-overall} the SpTrSV runtimes into performance profiles~\cite{dolan2002benchmarking}. We have generated a performance profile for each data set and processor architecture, cf.\@~\S\ref{sec:experiment-setup}. In each of these performance profiles, we took the best SpTrSV runtime for each core count and matrix combination. For a given algorithm, we then computed the ratio of all SpTrSV from that algorithm that are with a given threshold of the respective best SpTrSV runtime. The resulting curve is then plotted with the x-axis representing the threshold and the y-axis representing the ratio.


The closer a curve of a given algorithm is to the top left corner, the better the algorithm is at performing parallel SpTrSV. We can clearly see that the new algorithm ElasticDivide performs exceptionally well on the ARM architecture. 
Similarly, on x86, ElasticDivide is the most performant, though it shares the top spot with GrowLocal. 
We explain this with the increased NUMA domains the ARM Kunpeng architecture has, favouring asynchronous compute.

SpMP is a strong contender on the SuiteSparse data set and performs well overall, but ultimately falls a bit short of ElasticDivide and GrowLocal.
HDagg struggles with the hard to parallelise SuiteSparse data set and only seems to do well on x86 Xeon architecture with smaller core count.

\subsection{Scaling}
\label{sec:scaling}

We investigate how the algorithms scale with the core count by graphing the geometric mean and interquartile range of the speed-up over serial execution. We do this in Figure~\ref{fig:scaling} for every pair of architecture and data set, cf.~\S\ref{sec:experiment-setup}.

We once more see that ElasticDivide outperforms all other algorithms on all data sets on the ARM architecture. 
It reaches a geometric mean speed-up of $20.00$ on the Metis data set on $64$ cores. 
This is 12\% higher than the speed-up of $17.93$ reached by the second best algorithm GrowLocal on the same data set (and architecture).

On the x86 architecture, GrowLocal and ElasticDivide compete closely for the best algorithm. However, as the core count increases, in particular on the SuiteSparse data set, SpMP continues to improve relative to GrowLocal and ElasticDivide and in some cases matches or improves upon their performance.

HDagg does not scale well with the number of cores except for the Metis data set as the nested dissecting reordering works well with the wavefront-gluing method that HDagg employs.

On all architectures, we see a drop in speed-ups at the $80$ core count. On the EPYC processor this coincides with a second socket, which is a likely contributor, but on the Kunpeng processor we already require the second socket at $64$ cores. Thus, we find the more likely explanation is that the parallelism which is exposed by the matrices and then picked up by the algorithms peters out around that core count, at least for a majority of matrices.
The reader may consult the Tables~\ref{table:florida-graph-table}, \ref{table:cholesky-graph-table}, and \ref{table:hdagg-metis-graph-table} for the average wavefront size of the matrices.

\subsection{Synchronisation}

\begin{table}[!htbp]
	\centering
	\begin{tblr}{
			colspec = {Q[l,m] |  Q[c,m] || Q[si={table-format=1.2, table-number-alignment=center},c] | Q[si={table-format=2.2, table-number-alignment=center},c] | Q[si={table-format=2.2, table-number-alignment=center},c]  },
			row{1} = {font=\bfseries, guard},
			column{1,2} = {font=\bfseries},
			rowhead = 1,
			rowfoot = 0,
		}
		Data& Cores & HDagg & GrowLocal & {ElasticDivide}  \\ \hline \hline
		\SetCell[r=4]{m}{Suite} & 16	& 1.34	& 16.02	& 6.74 \\ \hline
		& 32	& 1.15	& 12.78	& 5.10 \\ \hline
		& 64	& 1.06	& 11.35	& 4.48 \\ \hline 
		& 128	& 1.07	& 10.69 	& 4.17 \\ \hline
		\SetCell[r=4]{m}{iChol} & 16	& 1.64	& 20.74	& 10.33 \\ \hline 
		& 32	& 1.52 	& 17.35	& 8.62 \\ \hline
		& 64	& 1.44	& 14.58	& 7.76 \\ \hline
		& 128	& 1.37	& 13.30	& 7.01 \\ \hline
		\SetCell[r=4]{m}{Metis} & 16	& 2.57	& 16.73	& 8.59 \\ \hline
		& 32	& 2.10	& 14.54	& 8.15 \\ \hline
		& 64	& 1.87	& 12.70	& 7.22 \\ \hline 
		& 128	& 1.74	& 11.38 	& 6.73
	\end{tblr}
	
	\caption{\label{table:barrier}%
		Geometric-mean reduction of the number of synchronisation barriers over the number of wavefronts for each algorithm.
	}
\end{table}
In Table~\ref{table:barrier}, we present the geometric-mean reduction of the number of synchronisation barriers (weak or strong) relative to the number of wavefronts of the matrix.
In other words, the reduction of synchronisation barriers compared to a wavefront scheduler.
The numbers we present are from all data sets over a sample of core counts.
We note that SpMP is missing in the table as it employs an asynchronous (point-to-point) method of compute.

We see that the wavefront-gluing technique of HDagg is more effective on the Metis nested dissection reordered data set and also at lower core counts. At higher core counts, in particular on the SuiteSparse data set, it struggles to glue together wavefronts. Under these settings, HDagg only becomes marginally better than a wavefront scheduler, which perhaps explains its lack of performance in the higher core count numbers.

Compared to the breadth-first search of wavefront-based schedulers like HDagg, the depth-first search approaches of GrowLocal and ElasticDivide are able to pack in more compute before having to synchronise. Albeit the reported reduction by ElasticDivide is about half the one by GrowLocal, one has to recall that the synchronisations by ElasticDivide are weak and since the staleness is $2$, cf.~\S\ref{sec:stale-synchronous-parallel}, its numbers should be multiplied by two. Having taken that into account, we see that ElasticDivide does slightly better than GrowLocal on the Metis data set and slightly worse on the SuiteSparse data set.

\subsection{Overlap of Synchronisation and Compute}
\label{sec:bsp-vs-ssp}

\begin{table}[!htbp]
	\centering
	\begin{tblr}{
			colspec = {Q[l,m] | Q[c,m] ||  Q[si={table-format=1.2, table-number-alignment=center},c] | Q[si={table-format=1.2, table-number-alignment=center},c] | Q[si={table-format=1.2, table-number-alignment=center},c] },
			row{1} = {font=\bfseries, guard},
			column{1, 2} = {font=\bfseries},
			rowhead = 1,
			rowfoot = 0,
		}
		Architecture & Cores & Suite & iChol & Metis\\ \hline \hline
		\SetCell[r=6]{m}{ARM Kunpeng} & 16 & 1.57 & 1.07 & 1.06\\ \hline
		& 32 & 1.95 & 1.15 & 1.13 \\ \hline
		& 48 & 2.17 & 1.35 & 1.21\\ \hline
		& 64 & 2.82 & 1.37 & 1.34\\ \hline
		& 80 & 3.23 & 1.76 & 1.48\\ \hline
		& 96 & 4.51 & 2.02 & 1.65\\ \hline
		\SetCell[r=8]{m}{x86 EPYC} & 16 & 1.23 & 1.08 & 1.07\\ \hline
		& 32 & 1.36 & 1.09 & 1.07\\ \hline
		& 48 & 1.47 & 1.12 & 1.06\\ \hline
		& 64 & 1.60 & 1.12 & 1.05\\ \hline
		& 80 & 1.84 & 1.13 & 1.03\\ \hline
		& 96 & 1.93 & 1.18 & 1.01\\ \hline
		& 112 & 2.01 & 1.16 & 0.96\\ \hline
		& 128 & 2.19 & 1.13 & 0.99\\ \hline
		\SetCell[r=4]{m}{x86 Xeon} & 11 & 1.11 & 1.05 & 1.03\\ \hline
		& 22 & 1.33 & 1.08 & 1.12\\ \hline
		& 33 & 1.34 & 1.13 & 1.12\\ \hline
		& 44 & 1.15 & 1.10 & 1.08
	\end{tblr}
	\caption{\label{table:bsp-vs-ssp}%
		Geometric-mean speed-ups of stale-synchronous-parallel execution over bulk-synchronous-parallel execution for all architectures, core counts, and data sets.
	}
\end{table}

In order to investigate the effect of overlapping synchronisation and compute, we execute the stale-synchronous-parallel schedule computed by ElasticDivide twice:
once using our stale-synchronous-parallel SpTrSV kernel, cf.\@~\S\ref{sec:kernel}, and once using a bulk-synchronous-parallel SpTrSV kernel, implemented using OpenMP~\cite{dagum1998openmp}.
We note that this results in a valid execution, cf.\@ Remark~\ref{rem:ssp-schedule-also-smaller-staleness}.
In Table~\ref{table:bsp-vs-ssp}, we display the relative speed-up that the stale-synchronous-parallel SpTrSV kernel achieves over the bulk-synchronous-parallel one.

The data shows a clear picture:
overlapping synchronisation and compute is beneficial.
In general, we see higher improvements for data sets whose matrices are harder to parallelise such as the SuiteSparse data set and lower numbers for easier to parallelise matrices which are present in the Metis nested dissection data set.
We also see that the costlier the synchronisation is, for example through higher non-uniform memory architectures, the larger the improvement is.
That is, we see larger speed-ups in ARM than x86 and larger speed-ups for higher core count.
There is one exception to the latter and that is the x86 EPYC architecture on the Metis data set.
This phenomenon requires further investigation, though we offer possible explanations:
the compiler produces a more efficient type of barrier for this architecture than our barrier, cf.\@~\S\ref{sec:barrier} and \cite{hoefler2005survey}, and/or the stale-synchronous-parallel schedule looks kind of like a bulk-synchronous-parallel schedule, that is every other superstep is almost empty.

\subsection{Amortisation}
\label{sec:amortisation}

\begin{figure*}[!htbp]
	\centering
	\includegraphics[width=0.99\linewidth]{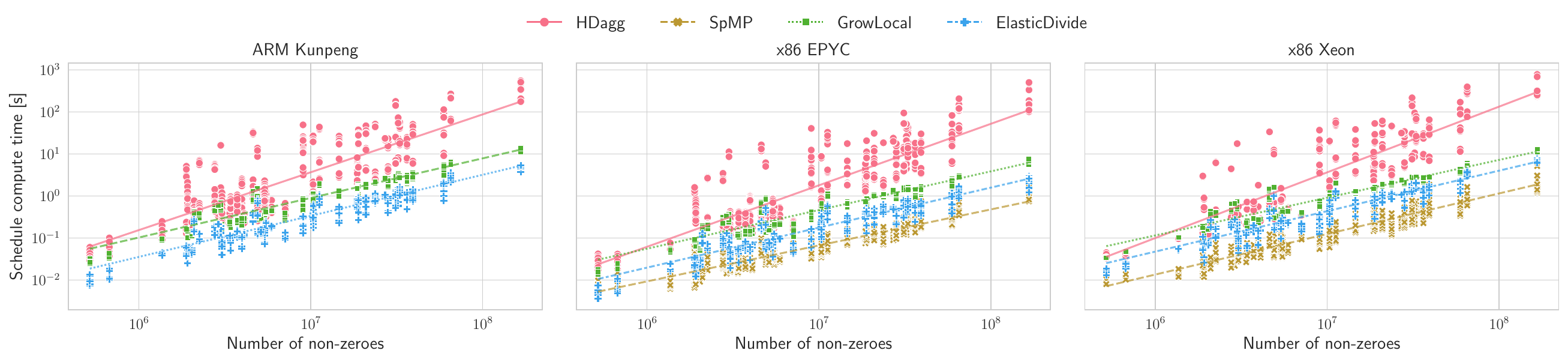}
	\caption{\label{fig:scheduling-time} Scheduling times of each algorithm plotted against number of non-zeroes of a matrix. The best $\ell^2$-fitted line of the shape $\log(y) = m \cdot \log(x) + c$ is also depicted.}
\end{figure*}

\begin{table*}[!htbp]
	\centering
	\begin{tblr}{
			colspec = {Q[l,m] | Q[l,m] ||  Q[si={table-format=4.2, table-number-alignment=center},c] | Q[si={table-format=1.2, table-number-alignment=center},c] | Q[si={table-format=2.2, table-number-alignment=center},c] | Q[si={table-format=2.2, table-number-alignment=center},c] },
			row{1, 2} = {font=\bfseries, guard},
			column{1, 2} = {font=\bfseries, guard},
			rowhead = 1,
			rowfoot = 0,
			cell{3-5}{4} = {guard},
		}
		\SetCell[r=2]{m}{Architecture} & \SetCell[r=2]{m}{Data} & \SetCell[c=4]{m}{Algorithm}
		\\ \hline
		& & HDagg & SpMP & GrowLocal & ElasticDivide
		\\ \hline \hline
		\SetCell[r=3]{m}{ARM Kunpeng} & Metis & 54.82 & --- & 33.59 & 13.60
		\\ \hline
		 & iChol & 211.01 & --- & 35.38 & 13.16
		\\ \hline
		 & Suite & 1763.03 & --- & 31.10 & 11.59
		\\ \hline
		\SetCell[r=3]{m}{x86 EPYC} & Metis & 85.59 & 5.41 & 52.91 & 21.32
		\\ \hline
		& iChol & 331.20 & 7.13 & 53.88 & 22.97
		\\ \hline
		& Suite & 2819.82 & 9.20 & 58.28 & 17.51
		\\ \hline
		\SetCell[r=3]{m}{x86 Xeon} & Metis & 59.52 & 5.88 & 43.79 & 21.60
		\\ \hline
		& iChol & 227.06 & 7.54 & 46.66 & 22.80
		\\ \hline
		& Suite & 1398.14 & 7.20 & 45.51 & 25.10
	\end{tblr}
	\caption{\label{table:amortisation}%
		Median amortisation costs of each algorithm on a full single socket for each architecture and data set.
	}
\end{table*}

An important aspect of the parallel SpTrSV schedules is the time it takes to generates them in the first place and how the quality of the schedule compares to the overhead to generate it in the first place. To this end, we look at how the scheduling time increases with the number of non-zeroes in the matrix. We display the results in Figure~\ref{fig:scheduling-time}.

We also compute the amortisation costs associated with a schedule~\cite{zarebavani2022hdagg, GrowLocalIPDPS, GrowLocalArXiv}. This is defined as the number of executions at which point it is more beneficial to compute a parallel schedule and then execute in parallel, opposed to always using serial execution:

\begin{equation}
	n_{\rm amortisation} =
	\begin{cases}
		\frac{t_{\rm schedule}}{\overline{t_{\rm serial}} - \overline{t_{\rm parallel}}}, & \text{if } \overline{t_{\rm serial}} > \overline{t_{\rm parallel}}, \\
		+\infty, & \text{otherwise},
	\end{cases}
\end{equation}
where the bar on top of the times denotes arithmetic mean. In Table~\ref{table:amortisation}, we record the median amortisation costs for each algorithm, data set, and architecture, cf.~\S\ref{sec:experiment-setup}. As the core count, we have taken the number of cores corresponding to a single socket.

It is apparent that SpMP is the fastest in scheduling time and that it scales linearly with the number of non-zeroes.
This goes to show the engineering efforts that went into the algorithm, in particular the efforts to the make the scheduling algorithm parallel in the first place.
ElasticDivide and GrowLocal generate their schedule using a serial algorithm and are therefore slower than SpMP.
Nevertheless, their runtime scales (provably) linearly with the number of non-zeroes.
HDagg is considerably slower and does not scale linearly.
Indeed, the slopes in Figure~\ref{fig:scheduling-time} of the best $\ell^2$-fitted lines of the shape $\log(y) = m \cdot \log(x) + c$ are between $1.38$ and $1.56$ for HDagg, whereas they are $\le 1.0$ for all other algorithms.
Albeit that ElasticDivide and GrowLocal follow a similar algorithm design, the distance between their $\ell^2$-fitted lines in Figure~\ref{fig:scheduling-time} indicates that ElasticDivide is twice as fast in computing a schedule.
We attribute this to a better implementation with improved data structures.
The amortisation cost table, Table~\ref{table:amortisation}, reflects the scheduling time data depicted in Figure~\ref{fig:scheduling-time}.

SpMP comes in the lowest with an amortisation cost of $5$-$10$ across x86 architectures.
ElasticDivide achieves low amortisation costs, ranging from $12$ to $25$ across all architectures and data sets.
GrowLocal has costs ranging from $31$ to $58$, performing best on the Metis data set and worst on the SuiteSparse data set.
HDagg has the largest amortisation costs, ranging from $55$-$86$ on the most favourable Metis data set all the way up to $2820$ on the SuiteSparse data set on x86 EPYC.

\section{Conclusion}
\label{sec:conclusion}

Synchronising many cores is expensive, more so in non-uniform memory architectures.
To combat this expense, one may reduce the number of synchronisations necessary through preprocessing the matrix whenever possible, e.g., nested dissecting, through algorithms that reduce the number of synchronisations to a minimum, and, as we have shown in this paper, through overlapping compute with synchronisation with execution models such as stale-synchronous-parallel.
The latter can improve performance easily by 10-30\% and when the number of synchronisations is large and the architecture is highly non-uniform by 100\% and in some cases up to 350\%, cf.\@~\S\ref{sec:bsp-vs-ssp}.

Our stale-synchronous-parallel scheduler, ElasticDivide, demonstrates that the fine-grained dependency graphs present in sparse triangular linear systems allow for the necessary flexibility to fit into a stale-synchronous-parallel schedule without (significant) increase in the number of synchronisation events when appropriately normalised.
This resulted in geometric-mean speed-ups of around 7-30\% against state-of-the-art scheduler GrowLocal on ARM Kunpeng, with even higher speed-ups for higher core and NUMA-domain count, cf.\@~\S\ref{sec:scaling}.

Albeit our presentation and evaluation was conducted on a variety of CPUs, our algorithm ElasticDivide and more generally our ideas and findings are transferable to other compute architectures.
For instance, ElasticDivide is directly applicable to solving sparse triangular systems on accelerated compute systems such as GPUs.
There it could be used to mask CPU to GPU and cross-GPU latencies with compute.



\FloatBarrier

\appendix
\section{Tables of matrices} \label{sec:tables-matrices}

The tables provided here give more details on the the matrices used in the experiments, cf.\@ \S\ref{sec:data-sets}, together with some basic statistics.

\begin{longtblr}[
	theme = ams-theme,
	caption = {Matrices used in the evaluation from the SuiteSparse Matrix Collection~\cite{davis2011university}. The average wavefront size has been rounded down.},
	label = {table:florida-graph-table},
]{
	colspec = { Q[l,m] | Q[r,m] | Q[r,m] | Q[r,m]  },
	row{1} = {font=\bfseries},
	rowhead = 1,
	rowfoot = 0,
}
Matrix & Size & \#Non-zeroes & Average wavefront size
\\ \hline \hline af\_0\_k101 & 503,625 & 9,027,150 & 74
\\ \hline af\_shell7 & 504,855 & 9,046,865 & 135
\\ \hline apache2 & 715,176 & 2,766,523 & 1,077
\\ \hline audikw\_1 & 943,695 & 39,297,771 & 203
\\ \hline bmw7st\_1 & 141,347 & 3,740,507 & 199
\\ \hline bmwcra\_1 & 148,770 & 5,396,386 & 204
\\ \hline bone010 & 986,703 & 36,326,514 & 470
\\ \hline boneS01 & 127,224 & 3,421,188 & 156
\\ \hline boneS10 & 914,898 & 28,191,660 & 386
\\ \hline Bump\_2911 & 2,911,419 & 65,320,659 & 283
\\ \hline bundle\_adj & 513,351 & 10,360,701 & 57,039
\\ \hline consph & 83,334 & 3,046,907 & 139
\\ \hline Dubcova3 & 146,689 & 1,891,669 & 44
\\ \hline ecology2 & 999,999 & 2,997,995 & 500
\\ \hline Emilia\_923 & 923,136 & 20,964,171 & 176
\\ \hline Fault\_639 & 638,802 & 14,626,683 & 143
\\ \hline Flan\_1565 & 1,564,794 & 59,485,419 & 200
\\ \hline G3\_circuit & 1,585,478 & 4,623,152 & 611
\\ \hline Geo\_1438 & 1,437,960 & 32,297,325 & 246
\\ \hline hood & 220,542 & 5,494,489 & 365
\\ \hline Hook\_1498 & 1,498,023 & 31,207,734 & 95
\\ \hline inline\_1 & 503,712 & 18,660,027 & 287
\\ \hline ldoor & 952,203 & 23,737,339 & 141
\\ \hline msdoor & 415,863 & 10,328,399 & 59
\\ \hline offshore & 259,789 & 2,251,231 & 75
\\ \hline parabolic\_fem & 525,825 & 2,100,225 & 75,117
\\ \hline PFlow\_742 & 742,793 & 18,940,627 & 118
\\ \hline Queen\_4147 & 4,147,110 & 166,823,197 & 342
\\ \hline s3dkt3m2 & 90,449 & 1,921,955 & 60
\\ \hline Serena & 1,391,349 & 32,961,525 & 298
\\ \hline shipsec1 & 140,874 & 3,977,139 & 67
\\ \hline StocF-1465 & 1,465,137 & 11,235,263 & 487
\\ \hline thermal2 & 1,228,045 & 4,904,179 & 991
\end{longtblr}

\begin{longtblr}[
theme = ams-theme,
caption = {Matrices used in the evaluation from SuiteSparse Matrix Collection~\cite{davis2011university}. These matrices were transformed using the incomplete Cholesky method with AMD reordering of Eigen~\cite{eigenweb}. The average wavefront size has been rounded down.},
label = {table:cholesky-graph-table},
]{ colspec = { Q[l,m] | Q[r,m] | Q[r,m] | Q[r,m]  }, row{1} = {font=\bfseries}, rowhead = 1, rowfoot = 0}
Matrix & Size & \#Non-zeroes & Average wavefront size
\\ \hline \hline af\_0\_k101\_iCh & 503,625 & 9,027,150 & 195
\\ \hline af\_shell7\_iCh & 504,855 & 9,046,865 & 668
\\ \hline apache2\_iCh & 715,176 & 2,766,523 & 79,464
\\ \hline audikw\_1\_iCh & 943,695 & 39,297,771 & 138
\\ \hline bmw7st\_1\_iCh & 141,347 & 3,740,507 & 340
\\ \hline bmwcra\_1\_iCh & 148,770 & 5,396,386 & 89
\\ \hline bone010\_iCh & 986,703 & 36,326,514 & 340
\\ \hline boneS01\_iCh & 127,224 & 3,421,188 & 245
\\ \hline boneS10\_iCh & 914,898 & 28,191,660 & 521
\\ \hline Bump\_2911\_iCh & 2,911,419 & 65,320,659 & 1,048
\\ \hline consph\_iCh & 83,334 & 3,046,907 & 78
\\ \hline Dubcova3\_iCh & 146,689 & 1,891,669 & 1,594
\\ \hline ecology2\_iCh & 999,999 & 2,997,995 & 142,857
\\ \hline Emilia\_923\_iCh & 923,136 & 20,964,171 & 511
\\ \hline Fault\_639\_iCh & 638,802 & 14,626,683 & 422
\\ \hline Flan\_1565\_iCh & 1,564,794 & 59,485,419 & 689
\\ \hline G3\_circuit\_iCh & 1,585,478 & 4,623,152 & 88,082
\\ \hline Geo\_1438\_iCh & 1,437,960 & 32,297,325 & 768
\\ \hline hood\_iCh & 220,542 & 5,494,489 & 1,050
\\ \hline Hook\_1498\_iCh & 1,498,023 & 31,207,734 & 649
\\ \hline inline\_1\_iCh & 503,712 & 18,660,027 & 679
\\ \hline ldoor\_iCh & 952,203 & 23,737,339 & 3,317
\\ \hline msdoor\_iCh & 415,863 & 10,328,399 & 956
\\ \hline offshore\_iCh & 259,789 & 2,251,231 & 1,114
\\ \hline parabolic\_fem\_iCh & 525,825 & 2,100,225 & 19,475
\\ \hline PFlow\_742\_iCh & 742,793 & 18,940,627 & 240
\\ \hline Queen\_4147\_iCh & 4,147,110 & 166,823,197 & 719
\\ \hline s3dkt3m2\_iCh & 90,449 & 1,921,955 & 104
\\ \hline Serena\_iCh & 1,391,349 & 32,961,525 & 940
\\ \hline shipsec1\_iCh & 140,874 & 3,977,139 & 259
\\ \hline StocF-1465\_iCh & 1,465,137 & 11,235,263 & 2,990
\\ \hline thermal2\_iCh & 1,228,045 & 4,904,179 & 47,232
\end{longtblr}

\begin{longtblr}[
theme = ams-theme,
caption = {Matrices used in the evaluation from the SuiteSparse Matrix Collection~\cite{davis2011university}. These matrices were symmetrically permuted using the fill-reducing method `METIS\textunderscore NodeND' of \cite{karypis1998fast}. The average wavefront size has been rounded down.},
label = {table:hdagg-metis-graph-table},
]{ colspec = { Q[l,m] | Q[r,m] | Q[r,m] | Q[r,m]  }, row{1} = {font=\bfseries}, rowhead = 1, rowfoot = 0}
Matrix & Size & \#Non-zeroes & Average wavefront size
\\ \hline \hline af\_0\_k101\_metis & 503,625 & 9,027,150 & 610
\\ \hline af\_shell10\_metis & 1,508,065 & 27,090,195 & 1,065
\\ \hline apache2\_metis & 715,176 & 2,766,523 & 47,678
\\ \hline audikw\_1\_metis & 943,695 & 39,297,771 & 1,734
\\ \hline bmwcra\_1\_metis & 148,770 & 5,396,386 & 473
\\ \hline bone010\_metis & 986,703 & 36,326,514 & 1,326
\\ \hline boneS10\_metis & 914,898 & 28,191,660 & 2,401
\\ \hline bundle\_adj\_metis & 513,351 & 10,360,701 & 11,407
\\ \hline cant\_metis & 62,451 & 2,034,917 & 333
\\ \hline consph\_metis & 83,334 & 3,046,907 & 247
\\ \hline crankseg\_2\_metis & 63,838 & 7,106,348 & 86
\\ \hline ecology2\_metis & 999,999 & 2,997,995 & 62,499
\\ \hline Emilia\_923\_metis & 923,136 & 20,964,171 & 2,107
\\ \hline Fault\_639\_metis & 638,802 & 14,626,683 & 1,458
\\ \hline Flan\_1565\_metis & 1,564,794 & 59,485,419 & 2,569
\\ \hline G3\_circuit\_metis & 1,585,478 & 4,623,152 & 93,263
\\ \hline Geo\_1438\_metis & 1,437,960 & 32,297,325 & 2,887
\\ \hline gyro\_metis & 17,361 & 519,260 & 88
\\ \hline hood\_metis & 220,542 & 5,494,489 & 984
\\ \hline Hook\_1498\_metis & 1,498,023 & 31,207,734 & 4,059
\\ \hline inline\_1\_metis & 503,712 & 18,660,027 & 1,549
\\ \hline ldoor\_metis & 952,203 & 23,737,339 & 4,858
\\ \hline m\_t1\_metis & 97,578 & 4,925,574 & 268
\\ \hline msdoor\_metis & 415,863 & 10,328,399 & 1,856
\\ \hline nasasrb\_metis & 54,870 & 1,366,097 & 287
\\ \hline PFlow\_742\_metis & 742,793 & 18,940,627 & 1,023
\\ \hline pwtk\_metis & 217,918 & 5,926,171 & 511
\\ \hline raefsky4\_metis & 19,779 & 674,195 & 111
\\ \hline ship\_003\_metis & 121,728 & 4,103,881 & 494
\\ \hline shipsec8\_metis & 114,919 & 3,384,159 & 456
\\ \hline StocF-1465\_metis & 1,465,137 & 11,235,263 & 11,446
\\ \hline thermal2\_metis & 1,228,045 & 4,904,179 & 45,483
\\ \hline tmt\_sym\_metis & 726,713 & 2,903,837 & 26,915
\\ \hline x104\_metis & 108,384 & 5,138,004 & 306
\end{longtblr}

\section{Flops/s}
\label{sec:gflops}

\begin{sidewaysfigure*}
	\centering
	\vspace{0.65\hsize}
	\includegraphics[width=0.9\hsize]{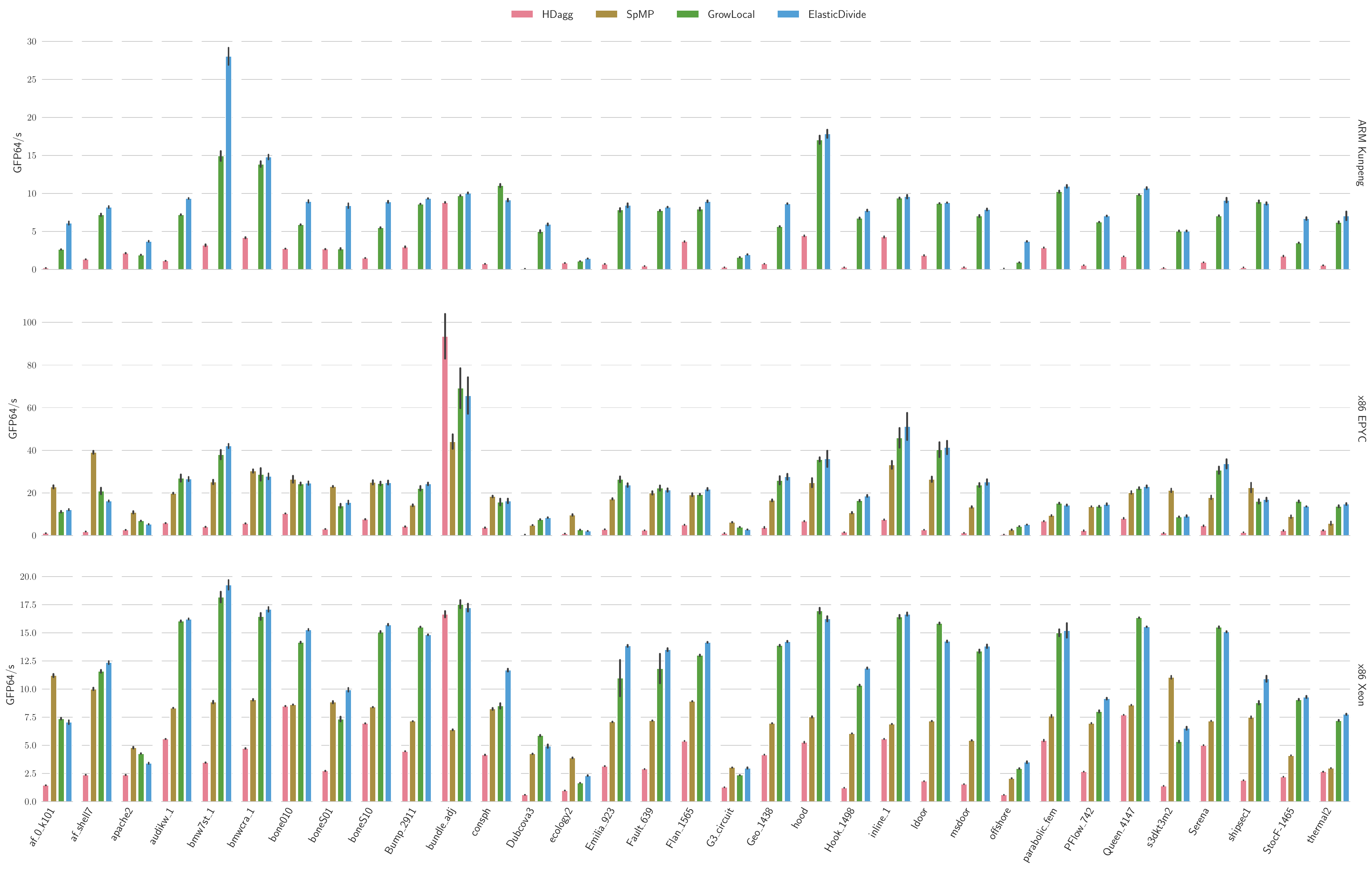}
	\caption{\label{fig:Gflop-SuiteSparse} Double precision floating point operations per second on the SuiteSparse data set using number of cores equal to one socket.}
\end{sidewaysfigure*}

\begin{sidewaysfigure*}
	\centering
	\vspace{0.65\hsize}
	\includegraphics[width=0.9\hsize]{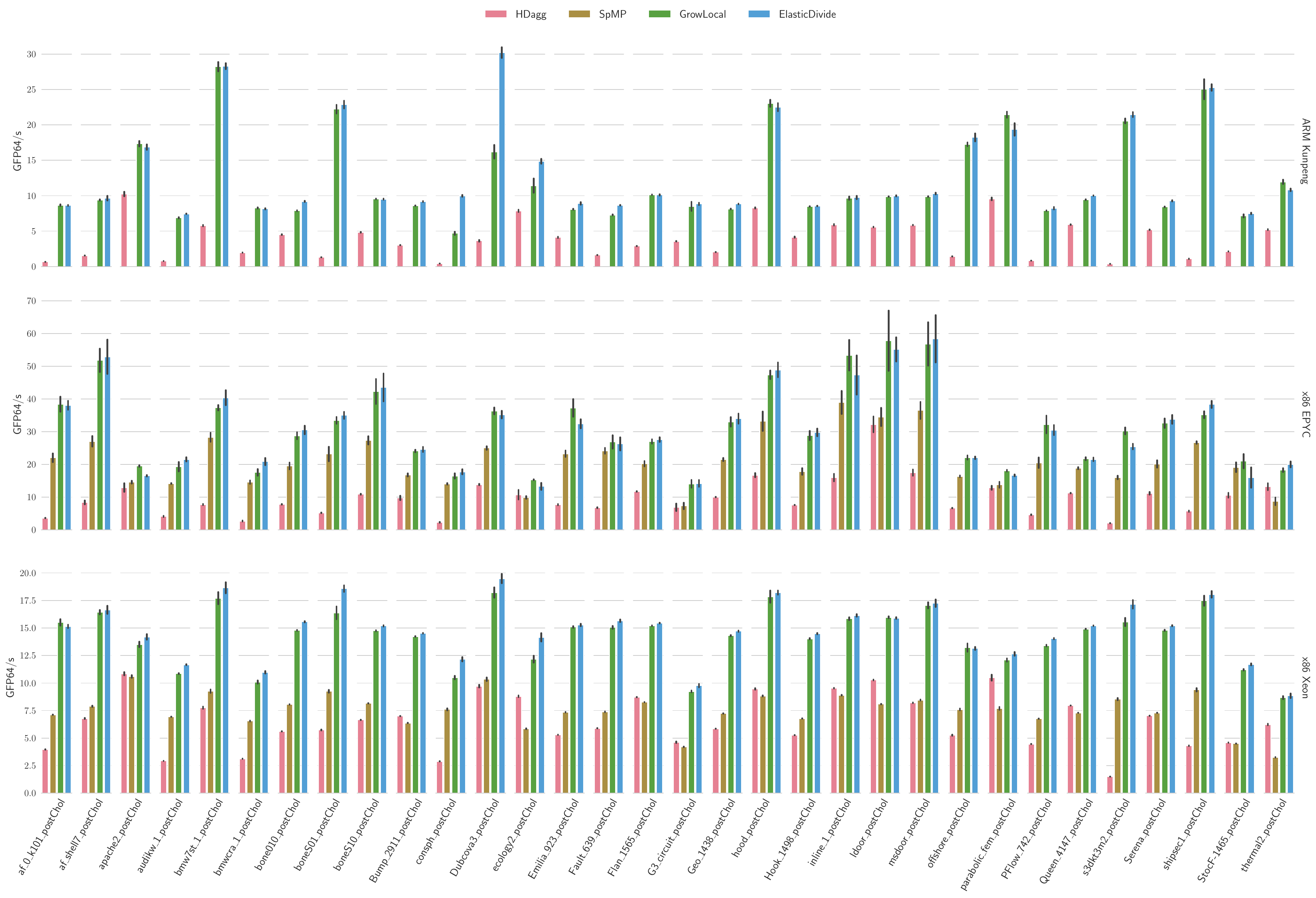}
	\caption{\label{fig:Gflop-iChol} Double precision floating point operations per second on the AMD reordered incomplete Cholesky data set using number of cores equal to one socket.}
\end{sidewaysfigure*}

\begin{sidewaysfigure*}
	\centering
	\vspace{0.65\hsize}
	\includegraphics[width=0.9\hsize]{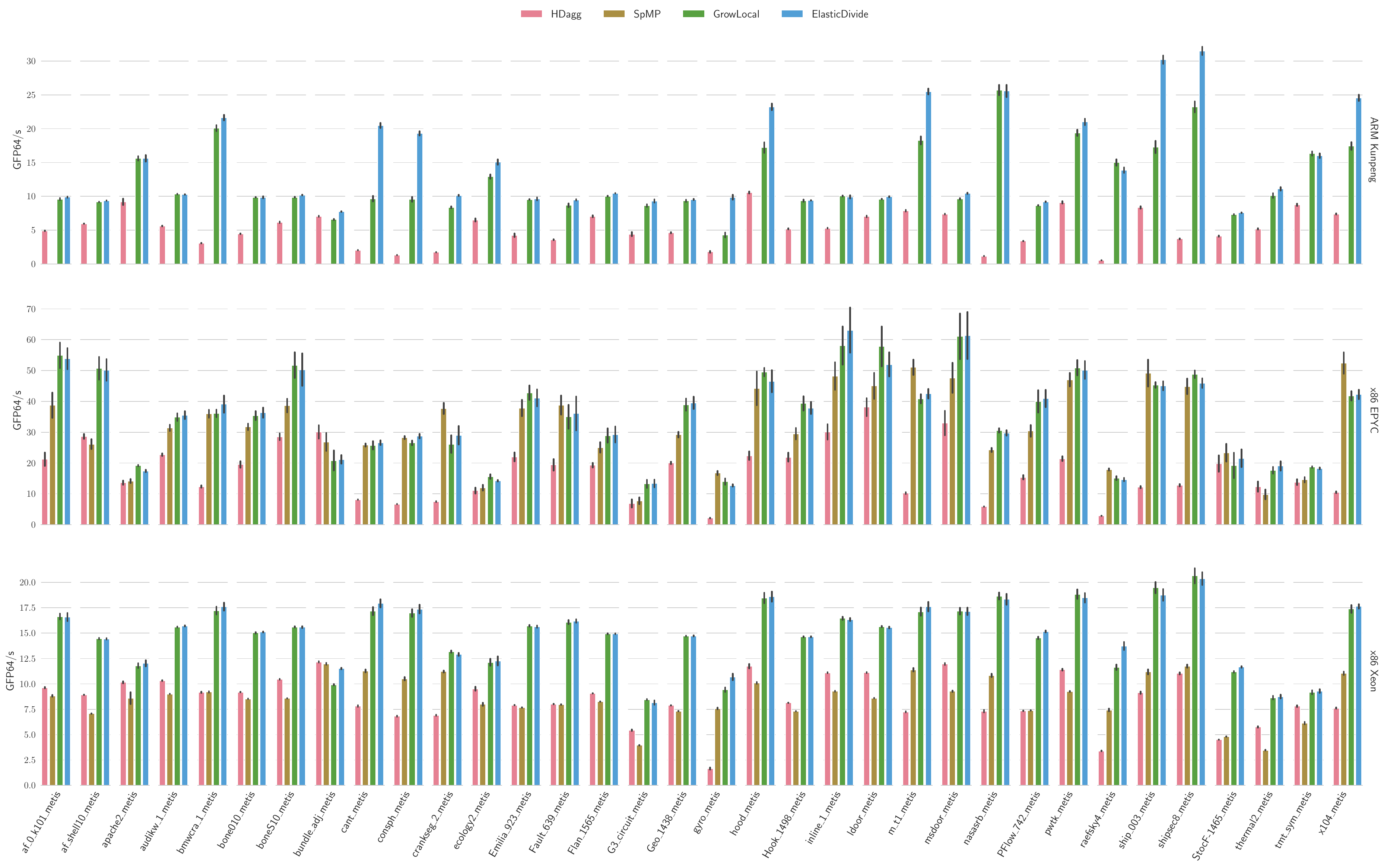}
	\caption{\label{fig:Gflop-Metis} Double precision floating point operations per second on the Metis data set using number of cores equal to one socket.}
\end{sidewaysfigure*}

In Figures~\ref{fig:Gflop-SuiteSparse}, \ref{fig:Gflop-iChol}, and \ref{fig:Gflop-Metis}, we depict the double precision floating point operations per second on each graph, data set, and architecture. The number of cores was chosen as a single socket. Error bars indicate the standard deviation of the measurements taken.

\FloatBarrier

\bibliographystyle{alpha}
\bibliography{references}

\end{document}